# Source Printer Classification using Printer Specific Local Texture Descriptor

Sharad Joshi and Nitin Khanna

## Abstract

The knowledge of source printer can help in printed text document authentication, copyright ownership, and provide important clues about the author of a fraudulent document along with his/her potential means and motives. Development of automated systems for classifying printed documents based on their source printer, using image processing techniques, is gaining lot of attention in multimedia forensics. Currently, state-of-the-art systems require that the font of letters present in test documents of unknown origin must be available in those used for training the classifier. In this work, we attempt to take the first step towards overcoming this limitation. Specifically, we introduce a novel printer specific local texture descriptor. The highlight of our technique is the use of encoding and regrouping strategy based on small linear-shaped structures composed of pixels having similar intensity and gradient. The results of experiments performed on two separate datasets show that: 1) on a publicly available dataset, the proposed method outperforms state-of-the-art algorithms for characters printed in the same font, and 2) on another dataset[1] having documents printed in four different fonts, the proposed method correctly classifies all test samples when sufficient training data is available in same font setup. In addition, it outperforms state-of-the-art methods for cross font experiments. Moreover, it reduces the confusion between the printers of same brand and model.

## Index Terms

Printer Classification, Local Texture Patterns, Local Binary Pattern, Forgery Detection, Printer Dataset.

## I. INTRODUCTION

The digital age is already upon us, and we have started reading and writing on digital devices in place of a physical piece of paper. However, the coexistence of printed and digital documents is ensured by factors such as ubiquitousness, ease of use, cost and security of documents. One of the important forensic aspects of a printed document (in this work, a single printed page is referred to as a printed document) is the knowledge related to the source of that document. The knowledge of printer used to print a document (source printer) can not only help in criminal investigations but also safeguard the use of paper for legal, administrative and other official records. Source attribution of printed documents using digital techniques has gained significant importance in recent times.

---

[1]Code and dataset will be made publicly available with published version of this paper.

Traditional methods use chemical or microscopic techniques which are time-consuming, costly, may even be intrusive and require an expert examiner. On the contrary, working on the scanned image of a printed document converts it into a classical pattern recognition problem involving feature extraction and classification [1], [2]. So, in digital techniques, only an office scanner and computer are required.

This area has received significant attention by researchers in the last decade [1]–[8]. These techniques involve the use of an extrinsically embedded signature or watermark before or during the printing process as well as detection of source printer based on intrinsic signatures. The, early methods based on intrinsic signatures relied on printed documents scanned at high resolutions which would be a costly and time-consuming affair. A relatively new intrinsic approach is based on geometric distortions induced by a printer. However, these either require the original soft copy of the printed document or a way to generate the reference soft copy from the printed document which would be troublesome and/or inaccurate in most real-life applications. There is another category of intrinsic signature which is based on texture analysis. Most of the successful methods in this category extract features from all occurrences of a specific letter type (such as 'e'), font type and font size. Then a classifier is trained to predict labels of the test data [1], [2], [6], [7], [9]. Thus, extracting information from all letters require training multiple classifiers, one for each letter (differing by letter type/font type/font size). Also, each of those classifiers would require enough amount of input letters for proper training. Further, extraction process for each of those letters such as use of optical character recognition (OCR), limits the classification process. An attempt was made recently to extract information from all letters by use of a single classifier [10]. The feature extraction pipeline of [10] gave promising results when letters printed using a specific font were used to train as well as test. However, in many practical scenarios, testing data might consist of only a particular font or language, not present in the training data.

Building on the approach in [10], this work presents a new feature descriptor for source classification of printed documents with the aim of addressing the cross font problem, a scenario where font of letters in test data is different from font of letters present in the train data. The printer's signature related to texture is a result of non-uniform and unintentional imperfections in toner developed on a printed page. The local texture features around a pixel in a small neighborhood would be directly dependent on the way this toner appears or spreads in that neighborhood. Thus, we introduce a new printer specific local texture descriptor (PSLTD) to capture textures on the scanned image of a printed document.

The proposed method makes no assumption about the type of the letter. Following are the major contributions of this paper:

- Introduction of a novel set of printer specific local binary pattern based features which can capture printer generated local texture patterns.

- Demonstration that encoding and regrouping strategy based on the orientation of intensity and gradient-based small linear structures makes PSLTD a very powerful discriminative feature.

- Evaluation of the proposed system comprehensively on the publicly available dataset [2], [7], showing that, given sufficient amount of train data, it outperforms existing methods. Also, it reduces confusion between printers of the same brand and model.

- Encouraging results on cross font dataset, outperforming state-of-the-art methods by large margins.

The remainder of this paper is organized as follows. Section II briefly describes existing intrinsic signature based techniques for classifying the source printer of printed text documents. The details of our proposed feature descriptor i.e. PSLTD have been mentioned in Section III. The proposed systems is described in Section IV. A series of experiments have been conducted to test the efficacy of the proposed method. Their description and results have been discussed in Section V. Finally, we present the conclusions that can be derived from this work along with the pending challenges for future work in Section VI.

## II. REVIEW OF EXISTING METHODS

Source printer classification problem has been researched extensively in the past decade [11]. The proposed solutions either try to extract an intrinsic signature (introduced due to the printing process) or some extrinsic signature intentionally embedded before or during the printing process [12]. Though extrinsic systems are promising, but they require modification of printing mechanism. In this work, we only focus on intrinsic signatures which are a result of imperfections in the printer parts [11]. Further, printers treat text and images differently. In this work, we derive a printer's signature from the text content. So, this section provides a review of only text-based techniques. More details of image-based methods are provided in [2] and [7].

One of the early signatures used for attributing the source printer was based on banding phenomenon. Banding refers to the appearance of light and dark lines perpendicular to the direction of the paper movement inside the printer [11]. The banding frequencies appearing on a page are used to attribute its source printer. Though this technique provided promising results, it requires the printed documents to be scanned at a very high resolution (2400 dpi).

In the recent past, texture based methods have gained significant interest as they have been shown to work reasonably well for text documents scanned at lower resolutions (600 dpi). This category was kick-started when Mikkilineni *et al.* [1] utilized gray-level co-occurrence matrices (GLCM) as a feature descriptor for all occurrences of the letter 'e'. In particular, 22 statistical features are estimated for GLCM corresponding to each occurrence of 'e'. The final labels for each occurrence of 'e' are predicted using a 5-nearest neighbors classifier which is followed by majority voting to obtain one source printer label for each page. Subsequently, there were variations in classification techniques like using support vector machine (SVM) [13], and Euclidean distances [14]. Following a different strategy, Kee and Farid [9] proposed a method based on the estimation of a printer profile. In their method, first, a reference character (i.e., a particular occurrence of letter 'e') is chosen. It is convolved with remaining characters to identify and extract similar characters. All extracted characters are aligned with respect to the reference character, and they are subjected to PCA. A profile consisting of mean character and top p eigenvectors is generated for each printer. Elkasrawi and Shafait [15] took a different path and presented a technique based on obtaining statistical features from the noise residual of each character. Influenced by the work of [16], Tsai *et al.* [6] proposed a combination of discrete wavelet transform (DWT) and GLCM based features. In particular, 22 GLCM and 12 DWT features are extracted from all occurrences of a specific Chinese character followed by classification using SVM. This method was extended in [17] by including more features obtained after applying a spatial filter, Wiener filter, and Gabor filter.

The problem of source printer attribution received a great push when a comprehensive dataset of English and Portuguese documents was made publicly available by Ferreira *et al.* [7]. They use features extracted from GLCM extended in multiple directions and scales. Also, they introduced a new feature descriptor. It is derived from convolutional texture gradient filter (CTGF) based on filtering textures with a specific gradient. The features are extracted from 'e' as well as rectangular portions of a page with sufficient printing material, termed as frames. The authors have recently extended this work to visualize important attribution features by directly highlighting them on the printed document under test [18]. Recently, a data-driven approach was proposed by Ferreira *et al.* [2] using six convolutional neural networks (CNN) trained in parallel. The input to these CNN are extracted and cropped versions of all occurrences of letters 'e', 'a' and their average and median residuals. The features obtained from all versions (original and filter residuals) derived out of each occurrences of letter 'e' on a printed document are concatenated and are used to train a one-vs-one SVM. The same is done with all versions of 'a'. Final printer label for each page is obtained by taking a majority voting on predicted labels corresponding to both 'e' and 'a'. In a more recent work, a single-classifier-based system was proposed in [10] which simultaneously uses information from all letters in a single classifier and does not need specific letters. This work uses connected component analysis followed by region separation and border removal. Next, local ternary pattern [19] based features are extracted from all components and subsequently, a post-extraction pooling is applied on a batch size of 40 feature vectors. Finally, SVM is used to predict the source printer labels of printed text documents. This work reported both group-of-character (GOC) level and page level accuracies. In contrast, a very recent method proposes a decision-fusion model based approach for source printer classification [20]. In particular, it applies feature selection approach on a variety of features including local binary pattern (LBP), GLCM and DWT. On the other hand, we propose a new feature descriptor (PSLTD) to classify the source printer.

Another interesting work is a text-independent approach to identify source printer proposed by Zhou *et al.* [8]. However, it requires specially designed and patented equipment to scan fine textures on printed pages. In sharp contrast, all the other methods discussed so far are designed to work with commonly used office scanners. Apart from texture-based techniques, a new category of methods has developed. It is based on printer specific geometric distortion [4], [21]. These methods rely on translational and rotational distortions of printed text relative to its reference soft copy [22]–[24]. In addition, some specific methods have been designed to classify printing technique i.e. laser, inkjet or photocopier [3], [25]–[30]. A more detailed literature review of related techniques has been discussed in [2] and [7]. In this work, we propose a printer specific local binary pattern based descriptor which encapsulates discriminating features from a printed text document scanned at a considerably lower resolution (600 dpi) using conventional office scanner. Also, the proposed method takes the first step towards font independent source printer identification.

## III. PRINTER SPECIFIC LOCAL TEXTURE DESCRIPTOR (PSLTD)

For source printer classification, a recent system based on local texture patterns outperformed other systems based on hand-crafted features [10] and achieved accuracies on par with a data-driven system [2]. Comparison of various local binary patterns indicated that local ternary pattern (LTrP) operator [19], coupled with Gabor filtering [31], outperforms other local patterns [10]. However, Gabor LTrP (GLTrP) was originally introduced specifically for context-based image retrieval [19]. Though GLTrP produces good results, peculiar nature of printing process and its associated noise-based texture patterns indicate that this problem will benefit from local texture patterns specifically developed for printer attribution. The primary hypothesis of this work is that toner-ink appearance or spread on a printed page is unique to a printer. However, differences in letters' shapes lead to variations in texture patterns estimated from the entire letters. So, first, a criterion for thresholding and quantization of intensity differences concerning the center pixel is defined to circumvent the effect of shape. Secondly, a particular elementary but generic configuration of pixels is identified to act as a reference for encoding and regrouping strategy. This is done to improve the discriminative power of local texture features. As the sizes of letters in question are quite small (approximately $50 \times 40$), so we extract features from a small patch of size $3 \times 3$ pixels. In such overlapping patches of size $3 \times 3$, we observed that there exist groups of pixels with similar intensity and gradient, having linear shapes (specific details of their selection are described later using Equations 4-16). So, for our PSLTD descriptor, we identify one pixel wide linear-shaped groups of pixels which are symmetrically located around the center pixel at all possible orientations i.e. horizontal ($0°$), vertical ($90°$), forward slant ($45°$) and backward slant ($135°$) as shown in Figure 1. Thus, the dimensions of each linear structure will be fixed at $3 \times 1$. The distribution of these linear-shaped structures for printers contained in a publicly available dataset [7] is depicted in Figure 2. This figure shows that distribution of linear structures having different orientations is similar in all printers, and their counts are large enough. Thus, this supports our proposition of using these structures to generate frequency histograms (whose number of bins equals the number of a particular type of orientation of linear structure) equivalent to the probability of occurrence of extracted texture patterns.

The design of our PSLTD will emphasize on preprocessing, thresholding and quantization, and encoding and regrouping stages of local binary feature extraction. The work-flow of PSLTD

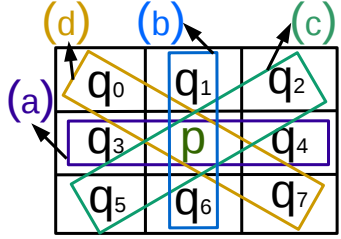

Fig. 1: Linear shaped structures with similar intensity and gradient. Four orientations have been used for encoding and regrouping: (a) horizontal ($0°$, depicted by purple color), (b) vertical ($90°$, blue color), (c) forward slant ($45°$, green color) and (d) backward slant ($135°$, brown color).

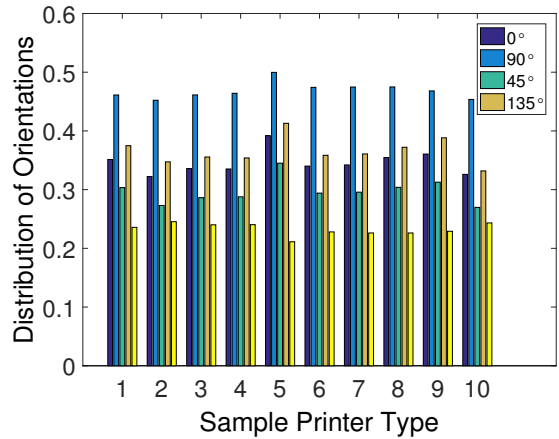

Fig. 2: Distribution of linear structures observed in overlapping image patches of size $3 \times 3$ (corresponding to Gabor filter's scale , $m = 0$). They are extracted from the bounding boxes of all letters contained in a sample document printed from all ten printers (Table I of [7] and [2]) of a publicly available dataset.

can be summarized as follows:

1) To begin with, a penta-pattern vector (PPV) is estimated from a contiguous $c \times c$ neighborhood around a center pixel $p$. If the neighborhood consists of $N$ neighbors, PPV will be of length $N$ such that each of its elements can take one of the five possible values. So, for $c = 3$, it will be an 8-element vector ($N = 8$). Let the intensity at $p$ be $I_p$. Then, at any pixel $p$, PPV is computed using the following relation;

$$
\begin{aligned}
PPV_{c,N}(p) &= \{p^0, p^1, ...., p^{N-1}\} \\
&= \{w_{PPV}(I_p - I_{q_0}), w_{PPV}(I_p - I_{q_1}), ..., w_{PPV}(I_p - I_{q_{N-1}})\}, \quad (1)
\end{aligned}
$$

where,

$$
w_{PPV}(x) = \begin{cases} 0, & |x| < T_0 \\ 1, & T_0 \leq x < T_1 \\ 2, & -T_1 < x \leq -T_0 \\ 3, & T_1 \leq x \\ 4, & x \leq -T_1 \end{cases} \tag{2}
$$

Here, $q_n$ represents $n^{th}$ neighbor of central pixel $p$, with n taking values from 0 to N-1. This setting tries to distinctly identify any undesirable noise introduced during the printing or scanning process using threshold $T_0$. Unlike LTP like patterns, the intensity difference is quantized into five levels using $-T_1$, $-T_0$, $T_0$, and $T_1$ as it provides more information about the non-uniform textures produced by toner ink. Specifically, the total number of possible patterns increases from $3^8$ to $5^8$ as compared to LTP. However, similar to LBP and LTP, this intensity difference based strategy ensures robustness to monotonic gray-scale variations.

2) Since binary patterns can be easily converted into compact features [19], the penta-pattern vector obtained in the previous step is converted into five binary pattern vectors (BPVs) as follows,

$$
BPV_{c,N,k}(p)|_{k=0,1,2,3,4} = \{\delta[p^0 - k], \delta[p^1 - k], ..., \delta[p^{N-1} - k]\}. \tag{3}
$$

Here, $\delta[.]$ denotes the Kronecker delta function and $p^n$ denotes the $n^{th}$ element of the PPV.

3) Next, irrespective of the PPV, for each overlapping patch in the letter image, linear shaped structures passing through the center pixel are identified. In general, each patch may have zero to four structures of varying orientations. As discussed, a group of pixels is defined as part of these linear structures if they have similar intensities and gradient directions. We choose the linear structures which satisfy both the constraints. For the similar intensity constraint case, an intensity line orientation vector ($\overrightarrow{EI}$) is estimated concerning the center pixel $p$ of each $3 \times 3$ overlapping patch. Let the location of $p$ be (x,y) and its intensity be $I(x,y)$, then $\overrightarrow{EI}$ is computed as follows:

$$
EI^0(x,y) = \begin{cases} 1, & |I(x,y) - I(x,y-1)| \leq T_0 \quad \text{and} \\ & |I(x,y) - I(x,y+1)| \leq T_0 \\ 0, & \text{else.} \end{cases} \tag{4}
$$

$$EI^1(x,y) = \begin{cases} 2, & |I(x,y) - I(x-1,y)| \leq T_0 \quad \text{and} \\ & |I(x,y) - I(x+1,y)| \leq T_0 \\ 0, & \text{else.} \end{cases} \tag{5}$$

$$EI^2(x,y) = \begin{cases} 3, & |I(x,y) - I(x-1,y+1)| \leq T_0 \quad \text{and} \\ & |I(x,y) - I(x+1,y-1)| \leq T_0 \\ 0, & \text{else.} \end{cases} \tag{6}$$

$$EI^3(x,y) = \begin{cases} 4, & |I(x,y) - I(x-1,y-1)| \leq T_0 \quad \text{and} \\ & |I(x,y) - I(x+1,y+1)| \leq T_0 \\ 0, & \text{else.} \end{cases} \tag{7}$$

$$EI^4(x,y) = \delta[EI^0(x,y) + EI^1(x,y) + EI^2(x,y) + EI^3(x,y)] \tag{8}$$

Here, $EI^l$ denotes the $l^{th}$ element of $\overrightarrow{EI}$. The threshold $T_0$ takes care of any undesirable noise. The $0^{th}$, $1^{st}$, $2^{nd}$ and $3^{rd}$ elements in $EI^l$ correspond to linear orientations of $0°$, $90°$, $45°$ and $135°$, respectively. If a particular orientation is present in the current image patch, then the corresponding element's value in $EI^l$ is set to '1' else it is set to '0' using Equations 4 to 8.

For gradient direction similarity, a multi-scale and multi-orientation Gabor filtered image is used as it has been shown to provide much better discriminatory gradient orientation information than the original image [19]. So, a Gabor filter of size $10 \times 10$ having 3 scales ($m = 0, 1, 2$) and 2 orientations ($0°$ and $90°$) is applied on the input letter image. This results in a Gabor filtered output $\Psi_{m\theta}(x,y)$ at each pixel. Now, for scale $m$, gradient line orientation vector ($\overrightarrow{EG_m}$) is computed similar to $\overrightarrow{EI}$ as follows:

$$EG_m^0(x,y) = \begin{cases} 1, & |G_m(x,y) - G_m(x,y-1)| \leq G_0 \quad \text{and} \\ & |G_m(x,y) - G_m(x,y+1)| \leq G_0 \\ 0, & \text{else.} \end{cases} \tag{9}$$

$$EG_m^1(x,y) = \begin{cases} 2, & |G_m(x,y) - G_m(x-1,y)| \leq G_0 \quad \text{and} \\ & |G_m(x,y) - G_m(x+1,y)| \leq G_0 \\ 0, & \text{else.} \end{cases} \tag{10}$$

$$EG_m^2(x,y) = \begin{cases} 3, & |G_m(x,y) - G_m(x-1,y+1)| \le G_0 \quad \text{and} \\ & |G_m(x,y) - G_m(x+1,y-1)| \le G_0 \\ 0, & \text{else.} \end{cases} \tag{11}$$

$$EG_m^3(x,y) = \begin{cases} 4, & |G_m(x,y) - G_m(x-1,y-1)| \le G_0 \quad \text{and} \\ & |G_m(x,y) - G_m(x+1,y+1)| \le G_0 \\ 0, & \text{else.} \end{cases} \tag{12}$$

$$EG_m^4(x,y) = \delta[EG_m^0(x,y) + EG_m^1(x,y) + EG_m^2(x,y) + EG_m^3(x,y)] \tag{13}$$

Here, $EG_m^l$ denotes the $l^{th}$ element of $\overrightarrow{EG_m}$. The gradient direction $G_m(x,y)$ at the location (x,y) is calculated from the Gabor filtered letter image as follows,

$$G_m(x,y) = tan^{-1}\left(\frac{|\Psi_{m0°}(x,y+1) - \Psi_{m0°}(x,y)|}{|\Psi_{m90°}(x-1,y) - \Psi_{m90°}(x,y)|}\right) \tag{14}$$

Thus, three $\overrightarrow{EG_m}$ are obtained at each pixel, corresponding to the three scales of Gabor filter ($m = 0, 1, 2$). The significance and values of each element in $\overrightarrow{EG_m}$ are similar to that for $\overrightarrow{EI}$. Now, for each scale $m$, the final $l^{th}$ line orientation ($E_m^l$) of a pixel is deduced by finding the linear structures which satisfy both the constraints (intensity and gradient direction similarity) as follows:

$$E_m^l(x,y) = \delta\left[EI^l(x,y) - EG_m^l(x,y)\right].EI^l(x,y), \ \forall \ 0 \le l \le 3 \tag{15}$$

$$E_m^4(x,y) = \delta[E_m^0(x,y) + E_m^1(x,y) + E_m^2(x,y) + E_m^3(x,y)] \tag{16}$$

The calculation of BPVs and combined line orientation vector ($E_m$) has been illustrated with an example in Figure 3.

4) An empirical analysis of some sample letters indicated that among the 255 possible types of patterns (corresponding to our 8-element BPV), uniform patterns [32] dominated in terms of occurrence. Uniform patterns represent fundamental micro-textures having a uniformity ($U$) value of two. A $U$ value of two implies that the maximum number of bit transitions from $0 \to 1$ or vice-versa in the 8-bit pattern considered circularly, is two. For any BPV, $U$ value can be computed using the dot product as follows:

$$U(BPV) = (BPV \oplus ROR(BPV, 1)).\overrightarrow{O} \tag{17}$$

Here, $ROR(BPV, 1)$ is the circular bit-wise right shifted version of BPV by 1 bit and $\overrightarrow{O}$ is an 8-bit vector with all bits set to '1'. Thus, we choose only the uniform patterns so that the total number of possible patterns is reduced to $58$. Next, all five types of BPVs for each pixel in the letter image are grouped. To model a printer's tendency of generating a particular orientation of the linear structure, a normalized histogram is constructed separately for each line orientation $E_m^l$. The number of bins in each of these histograms is fixed at $59$. Out of these, $58$ bins consist of normalized counts of uniform patterns and one bin has the normalized count of all the non-uniform patterns. Let $d$ denote a decimal code mapped to each of these bins i.e. $0 \leq d < 59$. This gives a feature vector of $25 \times 59 = 1475$ dimensions. It is computed as follows:

$$\vec{F}_{m,k,d,l} = \sum_{x=1}^{X} \sum_{y=1}^{Y} \delta \left[ \sum_{n=0}^{N-1} \left[ 2^n \left( BPV_{c,N,k} \left( p^n \right) \right) \right] - d \right] \times \delta \left[ E_m^l(x, y) - 1 \right],$$

$$\forall\ 0 \leq m \leq 2,\ 0 \leq k \leq 4,\ 0 \leq d < 59 \text{ and } 0 \leq l \leq 4 \qquad (18)$$

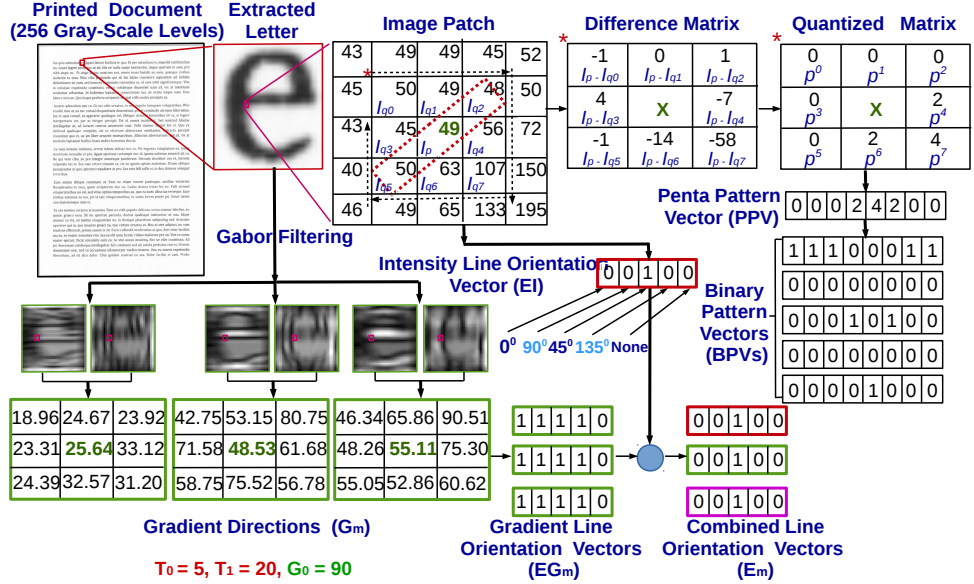

Fig. 3: Illustrative example for estimating binary pattern vectors (BPVs) and combined line orientation vectors ($E_m$).

Here, $l$ represents the corresponding line orientation ($0 \leq l \leq 4$). Further, the feature vectors obtained for each scale are concatenated into $1475 \times 3 = 4425$ patterns.

5) The magnitude component, when used with the sign of intensity differences, has been shown to improve discrimination power [33]. So, it is also included as a feature. It is defined as,

$$M_m(p) = \{u[Mag_m(q_0) - Mag_m(p)], ..., u[Mag_m(q_{N-1}) - Mag_m(p)]\}, \qquad (19)$$

where, $u[.]$ is unit step function and the gradient magnitude $Mag_m(.)$ is estimated as

$$Mag_m(x, y) = \sqrt{(|\Psi_{m0°}(x, y+1) - \Psi_{m0°}(x, y)|)^2 + (|\Psi_{m90°}(x, y+1) - \Psi_{m90°}(x, y)|)^2}.$$

6) For our printer specific local texture descriptor (PSLTD) we estimate two more sets of features. The first set of features ($F_1$) is estimated by considering only intensity constraint for encoding and regrouping strategy. For this purpose, $E_m^l$ is replaced by $EI^l$ in Equation 18. Similarly, for the second set of features ($F_2$), we consider only the constraint of gradient direction. This is done by replacing $E_m^l$ with $EG_m^l$ in Equation 18. In both of these cases, rest of the algorithm remains the same. Further, let the feature vector obtained from Equation 18 be denoted as $F_3$. Then, the proposed PSLTD is the concatenation of $F_1$, $F_2$ and $F_3$. Note here that computation of $F_1$ does not make use of Gabor filtering unlike $F_2$ and $F_3$. But, the magnitude component based patterns use Gabor filtered images of all three scales. So, $F_1$ will be of only 1652 dimensions ($1475 + 59 \times 3$). The final feature vector i.e. PSLTD comprises of $4602 \times 2 + 1475 + 59 \times 3 = 10856$ dimensions.

## IV. Printer Attribution Using PSLTD

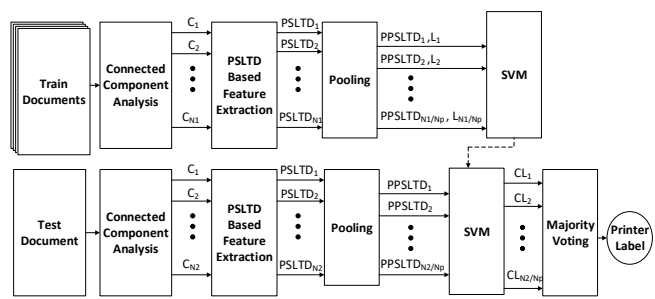

Fig. 4: Overall pipeline of the proposed approach. The proposed printer specific local texture descriptor (PSLTD) is extracted from each connected component (C) of train and test document. SVM is trained by pooled PSLTDs (PPSLTD) and corresponding printer labels for each character (L). The same model is used to predict labels for connected components (CL) obtained from test document. Finally, a majority vote on all CLs gives printer label for the test page.

In this section, we discuss all the steps involved in the proposed source printer attribution method (Figure 4). The hard copies of all printed documents are first scanned using a reference scanner. The user is free to choose the scanner as long as it is kept uniform across all printed documents. These scanned images of printed documents act as input to the proposed algorithm.

The proposed algorithm is composed of three major steps; 1) Connected component extraction, 2) PSLTD based feature extraction, and 3) classification using SVM. In the first step, we extract all connected components from a given text document using connected component analysis. We separate out spurious components by their area, width, and height. In particular, components larger than four times the median of areas of all components on a page are removed. Similarly, components smaller than 0.5 times the median of areas of all components on a page are also removed. This strategy works fine with the dataset used in [10] which does not contain any variation in font type and size across a page. But the dataset introduced in [7] contains printed Wikipedia pages which have drastically varying font sizes as well as font types. So, for experiments on this dataset using all letters, we further remove components with width and height smaller than 15 and 30 pixels, respectively. Besides, components with width and height larger than 90 and 100 pixels, respectively, are also removed. Though there can be an improvement in the proposed spurious component removal technique, nonetheless, it works fine for this research work.

After extracting letter images, we compute a PSLTD from each letter image using the procedure described in Section III. Now, due to the complex electro-mechanical parts and circuitry involved in the printing process, the occurrences of the same letter on a document printed in one go by a single printer may have different intensity distribution [34]. So, to reduce the variance among the PSLTDs extracted from the same page, we apply post extraction pooling (PoEP) [10] on all the feature vectors. Such a pooling averages feature vectors obtained from a group of $N_p$ letters.

Steps (1) and (2) are repeated for all train and test documents. Then, an SVM (LIBSVM implementation [35]) is trained using pooled feature vectors obtained from letters of train documents. In particular, a radial basis function kernel is used while the model parameters (c - cost and g - gamma) are selected using the default grid search optimization of LIBSVM. The grid for c (gamma) is from -5 and 15 (15 and 3) while the step size in the grid is fixed at 2 for both parameters. Next, pooled feature vectors extracted from letters printed on test documents are input to the trained SVM model which predicts the printer labels for each group of letters. Finally, a majority voting on predicted printer labels of all groups of letters present in the test document predicts the whole document's printer label. Also, before using feature vectors for training the SVM, we selectively filter out features/dimensions (from all the feature vectors) with almost zero information. Specifically, we remove features/dimensions which satisfy following constraints: (i) have non-zero values for less than 1% of the total number of training feature vectors; and (ii) their variance across all training feature vectors is less than $1 \times 10^{-9}$. Features/dimensions to be removed are decided based on training data and exact same features/dimensions are also removed from the testing data.

## V. Experimental Evaluation

We evaluate the performance of our PSLTD based method against state-of-the-art descriptors using a series of experiments. In particular, we compare against methods based on GLCM [1], multi-directional GLCM (*GLCM_MD*) [7], multi-directional multi-scale GLCM (*GLCM_MD_MS*) [7], *CTGF-GLCM-MD-MS$_e$* [7] and *CC-RS-LTrP-PoEP* [10]. In addition, comparison is done against the state-of-the-art data-driven system [2] denoted by *CNN-$\{S^{raw}, S^{med}, S^{avg}\}_{a,e}$*. Results obtained with the same font as well as cross font experiments are discussed in detail.

### A. Datasets and Experimental Setup

The most comprehensive publicly available dataset [7] (DB1) and another existing dataset, created by us [10] (DB2), are used to evaluate the performance of different methods. DB1 consists of 1184 Wikipedia pages printed from 10 printers (Table 1 in [2] and [7]) including a pair of printers with the same make and model. This dataset is available in the form of images of printed documents scanned at 600 dpi via a reference scanner (Plustek SO PL2546). The pages have characters comprising mixed font types and sizes, but their exact distribution is difficult to analyze. Also, some letters in italics and bold are spread randomly over different pages. On the other hand, the dataset DB2 consists of 720 text pages printed from 18 printers (Table I in supplementary material) with three printers having the same make and model. It comprises of pages printed at printer's default settings with four kinds of font types in the English language, Cambria (25 pages), Arial (5 pages), Times New Roman (5 pages) and Comic Sans (5 pages). Unlike DB1, the text in DB2 is randomly generated. Also, on any particular page letters are of only a single font type and size without any letters in bold and italics (Figure 1 in supplementary material shows one image for each type of font). So, DB2 is suitable for analyzing cross font scenarios. This dataset has been scanned at 600. For this work, we report all results on documents scanned at 600 dpi. In all the experiments, train and test sets are chosen such that they are disjoint. The value of pooling parameter $N_P$ is empirically chosen such that it acts as a hyperparameter of the proposed algorithm. It is kept small for lesser number of training pages and vice versa.

TABLE I: Comparison of average page-level classification accuracies of the proposed method against state-of-the-art methods on dataset DB1 [7] using $5 \times 2$ cross validation on the train and test folds provided by [2].

| Method | Accuracy (in %) |
|---|---|
| **GLCM$_e$** [1] | 77.87 |
| **GLCM_MD$_e$** [7] | 91.08 |
| **GLCM_MD_MS$_e$** [7] | 94.30 |
| **CTGF-GLCM-MD-MS$_e$** [2] | 96.26 |
| **CNN-$\{S^{raw}\}_e$** [2] | 96.13 |
| **CNN-$\{S^{raw}, S^{med}, S^{avg}\}_{a,e}$** [2] | 97.33 |
| **CC-RS-LTrP-PoEP$_e$** [10] | 97.12 |
| **PSLTD$_e$** | **98.92** |

Also, values of intensity and gradient orientation thresholds have been chosen empirically and kept fixed as $T_0 = 20$, $T_1 = 80$ and $G_0 = 90$ for DB1 which has been scanned using 256 grayscale levels. On the other hand, $T_0$ and $T_1$ are set to 13000 and 50000, respectively for DB2 since it has been scanned using 65536 grayscale levels. All experiments on the existing hand-crafted feature based methods are performed using the settings and parameters suggested by their authors. For experiments using data-driven based approach [2] on DB2, the validation loss and accuracy does not converge using an initial learning rate of 0.001. So, it was updated to 0.0001 for all the experiments.

### B. Experiments on DB1

First, we evaluate the performance of our proposed PSLTD based method on DB1 [7] which is publicly available. This dataset contains characters of mixed font types on the same page so we perform (i) letter specific (using all occurrences of letter 'e') evaluation similar to [2] and [7] and (ii) universal letter evaluation (using all connected components as suggested in [10]). Further, for a fair comparison, we use the same train and test folds as used by authors in [2] and [7] for reporting their results based on $5 \times 2$ cross-validation. Each fold has approximately 592 pages each for training as well as testing.

*1) Letter Specific Evaluation:* For the first set of experiments, we use all occurrences of the letter 'e' from each page instead of all the connected components. Rest of the pipeline remains the same as in Figure 4. The pooling parameter $N_p$ is chosen to be equal to the total number of occurrences of 'e' on a given printed document (as suggested in [10]). Under these settings, the proposed method (denoted as PSLTD$_e$) achieves an average page-level classification accuracy of 98.92% with a standard deviation of $\pm 0.43$ which is a gain of 1.8% over the best reported accuracy only on e's, using *CC-RS-LTrP-PoEP$_e$* [10] (Table I). In addition, the state-of-the-art CNN based method (*CNN-$\{S^{raw}, S^{med}, S^{avg}\}_{a,e}$* [2]) gives an average page-level classification accuracy of 97.33% using all occurrences of 'e' and 'a' images along with their median and average residuals. Thus, the proposed PSLTD$_e$ outperforms existing methods by correctly classifying, on an average, about 9 more pages out of 592 test pages. Moreover, the diagonal entries of the mean confusion matrix (i.e., the % of samples correctly classified per class) reveal that unlike state-of-the-art methods, the proposed method can significantly reduce the confusion between printers of same brand and model denoted by H225A and H225B (Table II). The mean confusion matrix obtained using the proposed method for all the ten folds is depicted in Table II, supplementary material.

*2) Universal Letter Evaluation (Mixed font experiment):* In this section, we discuss results using all connected components on a page as suggested in [10]. Under this setting, all connected components extracted from a particular page are used. Due to the presence of a variety of font types on a page, this setting can also be considered as a mixed font setting (i.e., letters of multiple font types are present in train data as well as test data). Similar to previous experiment, we fix $N_p$ equal to the number of extracted components on a given page. Further, the previous setting of using all occurrences of 'e' achieves a very high accuracy with current train-test split of DB1. Consequently, average page-level classification accuracy does not improve significantly when all the letters are used for training. In particular, the proposed method using all letters (denoted as PSLTD$_{All}$) provide an average page-level classification accuracy of 99.27% over ten folds with a standard deviation of 0.51. Thus, the proposed PSLTD$_{All}$ outperforms existing methods by correctly classifying, on an average, about 11 more pages out of 592 test pages.

Moreover, we also analyzed the proposed method with $N_p = 20$. This setting ensures that we have printer labels predicted for small groups of letters consisting of different types of printed letters. The proposed method gives an average group-of-letters level classification accuracy of 96.61% with a standard deviation of 0.41 when tested on about 64,000 groups of 20 letters each. Thus paving the way for potential intra-page tampering detection of certain types, for example, printing in vacant spaces and print, paste and copy forgeries [36] in such a mixed font scenario.

TABLE II: Page-level confusion matrix for $CNN\text{-}\{S^{raw}, S^{med}, S^{avg}\}_{a,e}$ [2], $CTGF\text{-}GLCM\text{-}MD\text{-}MS_e$ [2] and the proposed PSLTD on the existing dataset [7] corresponding to $5 \times 2$ cross-validation folds of [2].

| | B4070 | C1150 | C3240 | C4370 | H1518 | H225A | H225B | LE260 | OC330 | SC315 |
|---|---|---|---|---|---|---|---|---|---|---|
| **PSLTD_e** | 100 | 100 | 99.38 | 99.16 | 97.53 | 97.03 | 97.22 | 100 | 100 | 99.83 |
| **CC-RS-LTrP-PoEP_e [10]** | 99.61 | 97.26 | 98.54 | 98.67 | 94.99 | 93.47 | 91.95 | 98.35 | 98.35 | 100 |
| **CTGF-GLCM-MD-MS_e [2]** | 98.67 | 99.28 | 97.83 | 98.50 | 86.83 | 96.98 | 87.10 | 98.66 | 100 | 99.17 |
| **CNN-$\{S^{raw}, S^{med}, S^{avg}\}_{a,e}$ [2]** | 99.50 | 99.48 | 98.83 | 100 | 89.17 | 93.10 | 93.45 | 99.50 | 100 | 100 |
| **Miss-classification** **(PSLTD_e)** | | | 0.62 (C4370) | 0.84 (SC315) | 2.47 (C1150) | 2.97 (H225B) | 2.78 (H225A) | | | 0.17 (C4370) |
| **Miss-classification** **(CC-RS-LTrP-PoEP_e [10])** | 0.39 (C3240) | 1.20 (B4070) | 0.96 (C4370) | 1.00 (SC315) | 3.81 (C1150) | 6.53 (H225B) | 7.47 (H225A) | 0.99 (C1150) | 0.99 (H225B) | |
| **Miss-classification** **(CTGF-GLCM-MD-MS_e [2])** | 1.00 (C3240) | 1.72 (C4370) | 2.17 (C4370) | 0.50 (C1150) | 10.33 (C1150) | 2.52 (H225B) | 12.90 (H225A) | 0.67 (C1150) | | 0.50 (C1150) |
| **Miss-classification** **(CNN-$\{S^{raw}, S^{med}, S^{avg}\}_{a,e}$ [2])** | 0.33 (C3240) | 0.52 (B4070) | 0.67 (C3240) | | 10.50 (C1150) | 6.90 (H225B) | 6.37 (H225A) | 0.33 (H225A) | | |

## C. Experiments on DB2

The proposed method is evaluated on DB2 for: (a) same font experiments (as suggested in [10]) i.e., printed text used to train as well as test comprise of the same font type; and (b) cross font experiments i.e., font types of printed text available for training is different from the one needed to be tested.

*1) Same font experiments:* We can extract features using all letters with DB2 even for same font experiments as discussed in Section V-A. This is possible as each page consists of letters printed in only a specific font type. First, we analyze the performance in the inter-model scenario (i.e., no two printers have the same brand and model) [10]. Specifically, we analyze the performance of our proposed PSLTD on sixteen printers (except LC10 and LC11 in Table I, supplementary material) using letters (extracted components) from twenty and five pages for training and testing, respectively. Under this scenario, PSLTD provides an average page-level classification accuracy of 100% over five iterations (with random train and test data) with $N_p$ equal to the number of extracted components (i.e., twenty feature vectors per class for training the SVM). However, its performance degrades when only five pages per printer are used for training because the number of feature vectors per printer reduces drastically. So, for fewer training pages, a lower value of $N_p$ is required. Based on initial experiments, we set $N_p = 20$ when the amount of train data is small. The results (second column in Table IV) indicate that with five pages of training data, our proposed PSLTD based method provides an average page-level classification accuracy of 98.75% when tested on letters extracted from twenty pages per printer. On the other hand, *CNN-*$\{S^{raw}, S^{med}, S^{avg}\}_{a,e}$ [2] and *CC-RS-LTrP-PoEP* [10] predict correct printer labels for all test pages. One possible reason for this could be that the size of our proposed PSLTD is thrice that of the feature vector used by *CC-RS-LTrP-PoEP* [10].

TABLE III: Comparison of mean confusion matrices (accuracy in %) for classifying 3 printers of same brand and model. Bold values correspond to our proposed method while values in bracket correspond to *CC-RS-LTrP-PoEP$_e$* [10]

| Predicted Class<br>True Class | LC9 | LC10 | LC11 |
|---|---|---|---|
| **LC9** | **100** [99.8] | **0** [0.1] | **0** [0.1] |
| **LC10** | **0** [0.1] | **97.1** [96.2] | **2.9** [3.7] |
| **LC11** | | **1.2** [3.8] | **98.8** [96.2] |

For intra-model scenario (i.e., multiple printers may be of same brand and model), we conduct experiments on all the eighteen printers of DB2. The proposed PSLTD achieves 100% classification accuracy when trained with twenty pages and tested on five pages (when feature vectors of all letters are pooled into one). Furthermore, we compare the performance of our proposed PSLTD with *CC-RS-LTrP-PoEP* [10] on printers of same brand and model [10]. The mean confusion matrices obtained using the train and test letters extracted (using connected component analysis) from five and twenty pages, respectively, over five iterations is depicted in Table III. Similar to DB1, the proposed method reduces the confusion among printers of the same brand and model in DB2 as compared to *CC-RS-LTrP-PoEP* [10].

*2) Cross font experiments:* These experiments analyze the efficacy of PSLTD in a cross-font setup, i.e., font types of letters in test data are absent in train data. Samples of extracted connected components used in this setup are depicted in Figure 1 of supplementary material along with their PSLTDs. We conduct this set of experiments under inter-model scenario. Due to the low amount of training data in this setup, $N_p$ is chosen as 20 for the proposed method. Table IV lists the cross font source printer classification performance of proposed PSLTD along with *CC-RS-LTrP-PoEP$_{All}$* [10] and *CNN-*$\{S^{raw}, S^{med}, S^{avg}\}_{a,e}$ [2] (best existing methods based on previous experimental results depicted in Table I). The results indicate that:

TABLE IV: Average classification accuracies (in %) for same and cross font experiments using sixteen printers of unique brand and model in DB2 [10] (i.e., except LC10 and LC11). SVM is trained using five pages printed in Cambria font (C). Second column depicts accuracies obtained when trained model is tested on rest twenty pages of Cambria. Third, fourth and fifth columns consist of accuracies obtained when test contains all the five pages of Arial (A), Times New Roman (T) and Comic Sans (S) font, respectively.

| Train Font | Cambria (C) | | | |
|---|---|---|---|---|
| **Test Font** | **C** | **A** | **T** | **S** |
| ***CNN-*** $\{S^{raw}, S^{med}, S^{avg}\}_{a,e}$ [2] | **100** | 17.50 | 36.25 | 2.50 |
| ***CC-RS-LTrP-PoEP*** $_e$ [10] | **100** | 6.25 | 12.50 | 18.75 |
| **Proposed** | 98.75 | **77.50** | **87.50** | **60.00** |

- Even though the existing handcrafted descriptors perform well with same font experiments, their performance degrades considerably with cross font experiments. In particular, *CC-RS-LTrP-PoEP*$_{All}$ provides an average page-level classification accuracy of 100% using an SVM model trained and tested using letters printed in Cambria font (Table VII in [10]). In sharp contrast, it provides an average page-level classification accuracy of less than 20% when the same SVM model is used to test letters printed in Arial (A), Times New Roman (T) and Comic Sans (S) fonts. Please note that *CC-RS-LTrP-PoEP*$_{All}$ also provides an average page-level classification accuracy of 100% with same font experiments using Arial, Times New Roman, and Comic Sans fonts (Table VII in [10] ). So, it can be inferred that *CC-RS-LTrP-PoEP*$_{All}$ extracts very low font independent information.

- The data-driven method *CNN-*$\{S^{raw}, S^{med}, S^{avg}\}_{a,e}$ also performs well when it is trained using Cambria font and tested on the same font (100% accuracy, $3^{rd}$ row - $2^{nd}$ column of Table IV). However, the classification accuracy provided by this method reduces sharply when fonts of letters to be examined are different from font available for training (maximum accuracy 36.25%). The intuitive reasoning for this might be that the CNN is trained only on letters of the same font. So, it does its best to learn features which could discriminate between letters of the same font printed from multiple printers. Thus, the clusters of sample data for each printer class are tightly bound with minimum inter class overlapping. On the other hand, using the same setup to predict the source printer of letters from a different font might lead to larger intra-class variation.

- Although the proposed PSLTD gives slightly lesser accuracy compared to existing state-of-art, 98.75% average accuracy for classifying letters printed in Cambria font when the training is also done using letters printed in Cambria font. Nonetheless, the performance of both the existing state-of-art and the proposed system saturates to 100% when large number of training pages (20 pages/printer used for training as in Section V-C1) are available and testing is done on letters of same font as used in training. But the proposed system outperforms the state-of-the-art descriptors by huge margins for all cross-font combinations, when font of letters used in testing is completely different from font of letters used in training.

Thus, the same font results in Table IV concur with published literature [2], [10] and clearly demonstrate that the existing methods do very well when the font of letters present in testing data is also present in training data. However, these methods are not suitable for the scenarios where font of letters in testing data is different from that in training data. Thus, there is a need for developing systems which can work in cross-font scenarios. The proposed system is a step in that direction and its performance is further evaluated with different font types (Table V). Here, letters extracted (using connected component analysis) from five pages of each font type are used for training as well as testing. Table V shows that although the accuracies for cross-font scenarios are much lesser than those achieved by existing and proposed systems in same font scenarios, still the proposed system is a very promising step in the direction of font independence and eventual goal of language independence. For example, for classifying 16 printers, a classifier trained only on 5 pages/printer of text printed in Arial, gives an average classification accuracy of 66.25% for text printed in Comic Sans, while accuracy of a random guess in such a set up will be 6.25%.

TABLE V: Average classification accuracies (in %) for cross font experiments using sixteen printers of unique brand and model in DB2 [10] (i.e., except LC10 and LC11). Train and test letters extracted from 5 pages each.

| Train Font | Arial (A) | | | Times New Roman (T) | | | Comic Sans (S) | | |
|---|---|---|---|---|---|---|---|---|---|
| Test Font | C | T | S | C | A | S | C | A | T |
| Accuracy | 84.25 | 77.50 | 66.25 | 82.50 | 62.50 | 63.75 | 51.00 | 62.50 | 51.25 |

# VI. Conclusion

This paper proposed a printer specific local texture descriptor (PSLTD) whose highlight is the encoding and regrouping strategy which enhances its discriminative power. Thus, improving the source printer classification performance when a single letter is used for training and testing (Section V-B1). Further, we also applied PSLTD on all individual letter images (extracted by connected component analysis) such that only a single classifier is required for all of them. We have examined the performance of PSLTD by conducting a series of extensive experiments carried out on two datasets; a publicly available dataset as well as another dataset created by us. The examination suggests that PSLTD outperforms state-of-the-art descriptors on dataset DB1 and correctly classifies all test samples when sufficient training data is available on DB2. In particular, for same font experiments, PSLTD reduces the confusion between printers of the same brand and model in both the datasets. Furthermore, PSLTD also achieves good group-of-letter-level classification accuracies (for classifying a group of 20 letters instead of the whole printed page). Thus paving the way for PSLTD's application in certain types of potential intra-page forgeries. On the other hand, with cross font experiments, even though accuracies are much lesser as compared to same font scenario. Still, at the cost of slight reduction in performance on same font data, PSLTD outperforms existing methods by huge margins, for the fonts which are not present in the training set (Table IV, decrease of 1.25% compared to a minimum increase of 51.25%). Thus, it acts as the first step towards achieving a font-independent descriptor. Future work may include further enhancement of the performance of the proposed PSLTD for cross font scenarios by optimizing various hyper-parameters related to it such as dimensionality of feature vector and number of letters used for grouping. Alternative to SVM and investigation of its performance in an open set identification scenario can also be explored.

## Acknowledgment

This material is based upon work partially supported by the Board of Research in Nuclear Sciences (BRNS), Department of Atomic Energy (DAE), Government of India under the project DAE-BRNS-ATC-34/14/45/2014-BRNS and Visvesvaraya PhD Scheme, Ministry of the Electronics & Information Technology (MeitY), Government of India. Any opinions, findings, and conclusions or recommendations expressed in this material are those of the author(s) and do not necessarily reflect the views of the funding agencies. Address all correspondence to Nitin Khanna at nitinkhanna@iitgn.ac.in.

TABLE I: Details of printers in dataset DB2 [10].

| Printer ID | Printer Brand | Printer Model | Printer Resolution (in dpi) | Printer Type |
|---|---|---|---|---|
| LB1 | Brothers | DCP 7065DN | $2400 \times 600$ | Laser |
| LC1 | Canon | D520 | $1200 \times 600$ | Laser |
| LC2 | Canon | I6570 | $2400 \times 600$ | Laser |
| LC3 | Canon | IR 5000 | $2400 \times 600$ | Laser |
| LC4 | Canon | IR 7095 | $1200 \times 600$ | Laser |
| LC5 | Canon | IR 8500 | $2400 \times 600$ | Laser |
| LC6 | Canon | LBP 2900B | $2400 \times 600$ | Laser |
| LC7 | Canon | LBP 5050 | $9600 \times 600$ | Laser |
| LC8 | Canon | MF 4320 | $600 \times 600$ | Laser |
| LC9 | Canon | MF 4820d | $600 \times 600$ | Laser |
| LC10 | Canon | MF 4820d | $600 \times 600$ | Laser |
| LC11 | Canon | MF 4820d | $600 \times 600$ | Laser |
| IE1 | Epson | L800 | $5760 \times 1440$ | Inkjet |
| IE2 | Epson | EL 360 | $1200 \times 600$ | Inkjet |
| LH1 | HP | 1020 | $600 \times 600$ | Laser |
| LH2 | HP | M1005 | $600 \times 600$ | Laser |
| LK1 | Konica Minolta | Bizhub 215 | $600 \times 600$ | Laser |
| LR1 | Ricoh | MP 5002 | $600 \times 600$ | Laser |

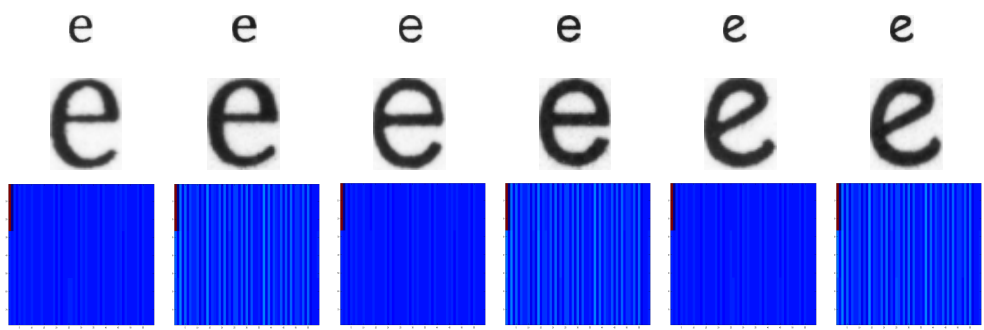

Fig. 1: Samples of letters 'e' (extracted using connected component analysis) alongwith their corresponding PSLTDs. First row corresponds to a pair of letters in Cambria font printed from two sample printers (LC4 & LC6) followed by a pair each of letters in Arial and Comic Sans font, respectively. Second row is a magnified version of first row. The respective PSLTDs of all letters are depicted in third row.

TABLE II: Confusion matrix (average page-level accuracies in %) for proposed method (Proposed$_e$) on publicly available dataset corresponding to $5 \times 2$ cross-validation folds reported in [2] and [7].

| True | Predicted | | | | | | | | | |
|---|---|---|---|---|---|---|---|---|---|---|
| | B4070 | C1150 | C3240 | C4370 | H1518 | H225A | H225B | LE260 | OC330 | SC315 |
| B4070 | 100.00 | | | | | | | | | |
| C1150 | | 100.00 | | | | | | | | |
| C3240 | | | 99.38 | 0.62 | | | | | | |
| C4370 | | | | 99.16 | | | | 0.84 | | |
| H1518 | 2.47 | | | | 97.53 | | | | | |
| H225A | | | | | | 97.03 | 2.97 | | | |
| H225B | | | | | | 2.78 | 97.22 | | | |
| LE260 | | | | | | | | 100.00 | | |
| OC330 | | | | | | | | | 100.00 | |
| SC315 | | | | 0.17 | | | | | | 99.83 |